\documentclass[acmsmall, screen, natbib=false]{acmart}

\renewcommand\footnotetextcopyrightpermission[1]{} 
\AtBeginDocument{%
  }

\usepackage{comment}
\usepackage{diagbox}
\usepackage{algorithm,algpseudocode, algorithmicx}
\algrenewcommand\alglinenumber[1]{\tiny #1:} 
\usepackage{setspace, etoolbox, caption}
\usepackage{graphicx}
\usepackage{epstopdf}
\usepackage{listings}
\usepackage{makecell,multirow}
\usepackage{url}
\usepackage{xcolor}
\usepackage{verbatim}
\usepackage{balance}
\usepackage{pifont}
\usepackage{enumitem}
\usepackage{adjustbox}
\usepackage{amsthm}
\usepackage{amsmath}
\usepackage{amsfonts}
\usepackage{textcomp,booktabs}
\usepackage[normalem]{ulem}
\usepackage{caption,subcaption}
\usepackage{tcolorbox}
\usepackage[T1]{fontenc}
\usepackage{threeparttable}
\usepackage{multicol}
\usepackage{flushend}

\newcommand{\eg}{\textit{e.g.,}}
\newcommand{\ie}{\textit{i.e.,}}
\newcommand\myparagraph[1]{
 \vspace*{3pt}
  \noindent \textit{\textbf{#1.}}\quad
}
\RequirePackage[
  datamodel=acmdatamodel,
  style=acmnumeric,
  ]{biblatex}
\begin{document}

\title{A Survey of Modern Compiler Fuzzing}

\author{Haoyang Ma}
\orcid{0000-0002-7411-9288}
\affiliation{
    \institution{Department of Computer Science and Engineering, The Hong Kong University of Science and Technology}
    \country{China}
}
\email{haoyang.ma@connect.ust.hk}

\begin{abstract}
Most software that runs on computers undergoes processing by compilers.
Since compilers constitute the fundamental infrastructure of software development, their correctness is paramount.
Over the years, researchers have invested in analyzing, understanding, and characterizing the bug features over mainstream compilers.
These studies have demonstrated that compilers' correctness requires greater research attention,
and they also pave the way for compiler fuzzing.
To improve compilers' correctness, researchers have proposed numerous compiler fuzzing techniques.
These techniques were initially developed for testing traditional compilers such as GCC/LLVM and have since been generalized to test various newly developed, domain-specific compilers, such as graphics shader compilers and deep learning (DL) compilers.
In this survey,
we provide a comprehensive summary of the research efforts for understanding and addressing compilers' defects.
Specifically, this survey mainly covers two aspects.
First, it covers researchers' investigation and expertise on compilers' bugs, such as their symptoms and root causes.
The compiler bug studies cover GCC/LLVM, JVM compilers, and DL compilers.
In addition, it covers
researchers' efforts in designing fuzzing techniques, including constructing test programs and designing test oracles.
Besides discussing the existing work, this survey outlines several open challenges and highlights research opportunities.
\end{abstract}

\maketitle
\fancyfoot{}
\thispagestyle{empty}

\section{Introduction}
\label{sec: intro}

Compilers are fundamental in software development. They translate a programming language into an executable form through single or several steps.
Virtually all codes, ranging from a single script to complicated software such as web browsers, necessitates compilers or compile-like tools, such as JavaScript engines, to execute.
Traditional compilers of mainstream program languages, such as GCC/LLVM for C/C++, have been widely used by millions of users for many years.
Nowadays, domain-specific languages (DSLs) have proliferated and have been widely adopted for specific tasks.
Compilers and compile-like tools for these DSLs, such as graphics shader compilers for shading languages (\eg GLSL) and DL compilers for DL models, have also been proposed to compile these DSLs and advance their efficiency.

Similar to other software, compilers are also prone to bugs~\cite{sunstudy, jvmlanguagebugstudy, qingchaoBugStudy, DuBugstudy, pyinterpreterstudy}. Compiler bugs may cause internal compiler errors, resulting in assertion failures or segmentation faults.
Additionally, compiler bugs may even implicitly induce the generation of incorrect binaries.
These misbehaviors can corrupt the development of software depending on these compilers.
For instance, a well-typed Java program containing some modern language features may be rejected by the compiler~\cite{jvmlanguagebugstudy}.
A well-formed DL model constructed by PyTorch~\cite{pytorch} can be mis-compiled by DL compilers~\cite{qingchaoBugStudy}.
These unexpected errors are confusing and misleading to Java programmers and DL engineers, making them wonder if the errors originate from their programs or models.
Besides, compiler bugs can also cause security issues.
For instance, David~\cite{david2018simple} reports a Microsoft Macro Assembler compiler bug. 
This bug is about misinterpreting negation: this assembler compiler neglects the negation during compilation and interprets the code \texttt{.if eax = !ebx} as \texttt{.if eax = ebx}. This bug offers hackers an opportunity of inserting backdoors into open-source projects.
For instance, hackers can access resources reserved for administrators by rewriting the checking for the user identity with negation.
Due to the severity of this critical bug, it has been transferred to Microsoft under CVE number \textit{CVE-2018-8232}.

The importance of compilers and the severe consequences caused by compilers' bugs illustrate the importance of compilers' correctness.
However, since compilers are complex, achieving correctness is nontrivial.
Take LLVM~\cite{LLVM} for instance. It contains several compilation steps, including 1) lexical analysis, 2) parsing and constructing the abstract syntax tree (AST), 3) generating LLVM intermediate representation (IR), 4) performing optimizations on the LLVM IR, 5) generating executable codes on the target platform.
Every compilation step is managed by a complex, sizeable, and ever-growing component.
For instance,
as new language features are added, the size of the lexical analysis and parsing components increases while new optimization strategies are incorporated into the optimization component.
TVM, a state-of-the-art DL compiler, is also synchronizing itself with the update of DL frameworks, integrating new optimization strategies, and incorporating new IRs~\cite{Relax}.

Numerous research efforts have been made to ensure the correctness of compilers.
For instance, Leory et al. investigate the formal verification of realistic compilers~\cite{Leroy_2009, leroy:hal-01238879, 10.1145/1538788.1538814, kastner:hal-01643290, kastner:hal-01399482} and propose CompCert, a verified compiler that is equipped with mathematical proof to ensure that the compiled executable code adheres precisely to the semantics of the source program.
Besides, researchers also apply 
translation validation~\cite{translationvalidation} to compilers~\cite{10.1145/358438.349314, 10.1145/349299.349314, 10.1145/3172871.3180078, 10.1145/3445814.3446751, 10.1145/3453483.3454030}.
In short, these techniques can scrutinize the semantics of the programs before and after compilation, verifying the preservation of semantics.
This survey does not encompass the aforementioned research efforts but rather concentrates solely on compiler fuzzing - a research avenue aimed at automatically constructing numerous test programs and utilizing them to identify compiler bugs. While compiler fuzzing cannot provide definitive proof or validation of the correctness of compilers, it can still be a valuable technique for reducing the number of defects in compilers. By identifying potential vulnerabilities or bugs in the compiler, fuzzing can help improve the overall quality and reliability of the compiler.

However, compiler fuzzing is challenging in four aspects. 
First, \textit{compiler fuzzing prefers valid (\ie syntax-compliant, semantics-compliant and semantics-unique) test programs.}
In practice, the grammar of modern programming languages is complex. 
Therefore, expertise is required to follow the grammar and write syntax-compliant programs.
Otherwise, the generated program would not pass the type checking in the front-end (\eg lexer and parser) of the compiler. 
Besides, being syntax-compliant is not enough sometimes. Some programming languages (\eg C) are constrained by undefined behaviors, meaning that the semantics of some syntax-compliant statements have more than one interpretation. To address this issue, the fuzzing technique should avoid generating undefined behaviors~\cite{Csmith, spirv-fuzz}.

Second, \textit{modern programming languages possess rich language features.}
The introduction of new language features is accompanied by an increase in grammar complexity.
The language feature is actually the production of grammar.
While constructing the source code for testing (\ie test program), exploring the production space of the grammar is not trivial. 
It requires well-designed search strategies, without which the generated test programs would not be diverse and thus cannot detect various bugs of different root causes.

Third, \textit{compilers' bugs may be implicit and not easily observable.}
The crash is obvious, specifying the compiler's internal error under test.
Therefore, utilizing crash as a test oracle to determine the compilation failure is common.
However, compilers' bugs do not always yield a crash.
For instance, 25.04\% of the collected 603 DL compiler bugs induce an incorrect result or a middle status without a crash~\cite{qingchaoBugStudy}.
These bugs manifest themselves quietly and implicitly, requiring more effective test oracles to be caught.


Motivated by these challenges, researchers have proposed 1) various program construction strategies to generate 
valid and diverse test programs, and 2) multiple test oracles to detect in-depth compiler bugs.

\myparagraph{Paper Organization}
Section \ref{sec: Paper Selection and Categorization} describes how we collect the papers.
Section \ref{sec: CompilerBugs} introduces the investigation of compiler bugs, outlining researchers' investigation of compiler bugs and showing that mainstream compilers are suffering from a large number of in-depth bugs with diverse symptoms and root causes.
Section \ref{sec: Compiler Fuzzing} is the core part of this paper, showing the development of evolvement of state-of-the-art compiler fuzzing techniques and approaches.
Section \ref{sec: Open Challenges and Research Opportunities} outlines the research challenges to be addressed.
\section{Paper Selection and Categorization}
\label{sec: Paper Selection and Categorization}

To carry out this survey, we gather 58 papers from international conferences and journals. These research efforts are for fuzzing modern compilers and compiler-like tools that are still actively under development and well-maintained.
Specifically, we collect fuzzing efforts for compilers of C/C++, Java, Java-like languages (\eg Kotlin, Scala, Groovy), OpenGL shader language (GLSL), and parallel programming languages (\eg OpenCL C/C++ and CUDA C/C++). Besides, this study also includes fuzzing techniques for DL compilers, DL inference engines, JVM implementations, JavaScript engines, and Simulink compilers. Though these tools are not compilers in the traditional sense, they do involve compilation as part of their workflow. 
For instance, 
JVM includes a just-in-time (JIT) compiler that dynamically compiles parts of the bytecode into machine code; 
JavaScript engine includes a bytecode compiler for compiling a JavaScript source code into JavaScript bytecode and a JIT compiler for interpreting the bytecode and compiling the bytecode into the executable;
DL compilers can compile DL models into different levels of IRs, and finally compile the IRs into executables on different platforms with the help of other compilers (\eg LLVM).

To collect relevant papers, we systematically search on Google Scholar with one of the following keywords: 
"compiler testing", "fuzz", "compiler defect", "compiler bug", "compiler vulnerabilit(y/ies)", "compiler fault".
And after filtering out papers with irrelevant topics, we obtain further papers by exploring the references in the papers already retrieved.
Since our topic is about "modern compiler fuzzing", 
we remove all compiler fuzzing work for old-fashioned compilers, such as the Fortran compiler, the Ada compiler, the Algol compiler, etc. These fuzzing efforts are well-outlined in a recent survey of compiler testing~\cite{JunjieSurvey}.
To ensure the quality of the papers collected, we mainly focus on papers published in top-tier conferences (Core Ranking A+ and A) and journals.
To increase the diversity of the papers, we also outline some inspiring ideas from papers published at non-top-tier conferences/journals/workshops/symposiums. Among the 58 papers, 12 papers are in this category.
Besides, we also include one master thesis~\cite{FuzzIL}, which demonstrates a great research effort in developing a fuzzer for JavaScript engines.

\begin{figure}[ht]
\centering
    \includegraphics[width=1.00\linewidth]{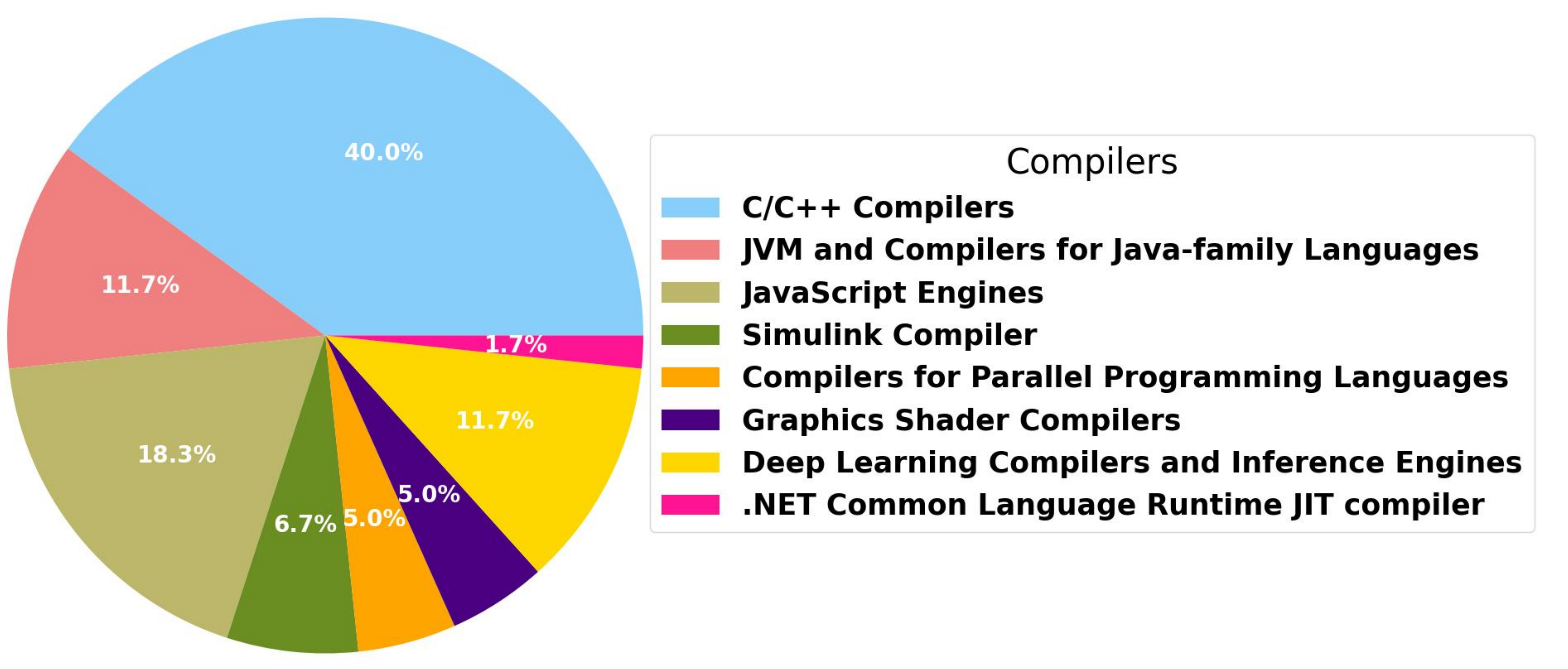}
    \caption{Paper Distribution on Each Compiler} 
    \label{figure:papersdistributionbycompiler}
\end{figure}

\begin{figure}[ht]
\centering
    \includegraphics[width=1.00\linewidth]{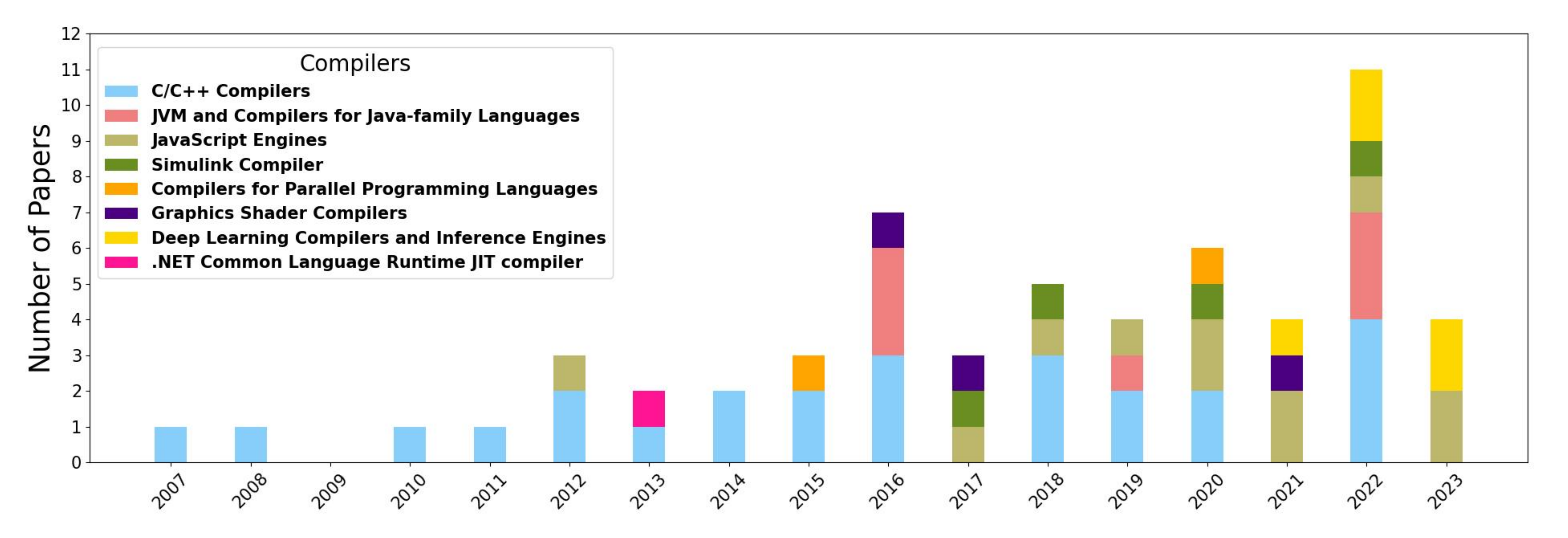}
    \caption{Compiler Fuzzing Papers Distribution by Years} 
    \label{figure:papersdistributionbyYear}
\end{figure}

Figure \ref{figure:papersdistributionbycompiler} illustrates the distribution of the 58 papers across different compilers or compiler-like tools.
Figure \ref{figure:papersdistributionbyYear} shows collected papers from 2007 to 2023. It also categorizes the published papers each year by their target compilers.
Out of these papers, 47 provide test program construction strategies, and 40 discuss their test oracle design.

\section{Compiler Bugs}
\label{sec: CompilerBugs}

As is mentioned in section \ref{sec: intro}, compiler bugs can cause ambiguity, confusion, and even severe security issues.
This section further outlines research efforts in collecting, analyzing, and understanding compiler bugs. 

\myparagraph{C/C++ Compilers}
GCC and LLVM are two mainstream and open-source compilers. They are mainly for compiling C++/C/Objective C++ code and generating executables.
Due to the popularity of these languages in both industry and academia, the bugs in them are worthy of attention.
Sun et al.~\cite{sunstudy} investigate their bugs and analyze them along four aspects: 
1) location of bugs; 
2) test cases, localization, and fixes of bugs;
3) duration of bugs; and
4) priorities of bugs. 
After exhaustively examining about 50K bugs and 30K bug fixes, they release several interesting findings:
1) The bug-revealing test cases are typically small. 80\% of them have fewer than 45 lines of code.
2) Most bug fixes only modify a single source file.
3) The average lifetime of GCC bugs is 200 days while that of LLVM bugs is 111 days.
4) C++ component is the most buggy.
Finding 1) and 2) imply that fuzzers for GCC/LLVM do not need to generate too many lines of code.
Finding 3) shows that the fixes for GCC/LLVM bugs are not easy. This finding shows that compiler testing is an important scientific research problem.
Finding 4) points the way for fuzzing GCC/LLVM.

Besides, Zhou et al.~\cite{ZHOU2021110884} exhaustively examine 8771 and 1564 optimization bugs of GCC and LLVM, respectively.
The analytical results reveal that
1) optimizations component is the second most bug-prone, after the C++ component,
2) the value range propagation optimization and the instruction combination optimization are the two most buggy optimizations in GCC and LLVM,
3) 57.21\% of the GCC optimization bugs and 61.38\% of the LLVM optimization bugs are mis-optimization bugs, which are the dominant bug types,
and
4) the average life span of bugs is over five months.

\myparagraph{Compilers of Java-family Languages}
Researchers also investigate typing-related compiler bugs~\cite{jvmlanguagebugstudy}. Specifically, they focus on studying bugs in the compiler front-ends of Java-family languages, including Java, Scala, Kotlin, and Groovy. 
With complicated type theories (\eg higher-kinded types~\cite{higher-kinded_types}, parametric polymorphism, and path-dependent types~\cite{Amin2016}) absorbed, type systems of these programming languages have become more and more complex.
And this growing complexity has led to difficulties in implementing the compiler front-ends, thus the growing number of front-end bugs.
In this work, researchers collect, randomly sample, and study typing-related bugs in four dimensions 1) symptoms, 2) root causes, 3) bug fixes, and 4) test case characteristics. Specifically, they categorize the sample 320 bugs into five symptoms (including \textit{Unexpected Compiler-time Error
}, \textit{Internal Compiler Error}, \textit{Unexpected Runtime Behavior}, \textit{Misleading Report}, and \textit{Compilation Performance Issue}) and five root causes (including \textit{Type-Related Bugs}, \textit{Semantic Analysis Bugs}, \textit{Resolution Bugs}, \textit{Bug Related to Error Handling \& Reporting}, and \textit{AST Transformation Bugs}). With case studies, this work clearly elaborates on the manifestation of each symptom and explanation of each root cause.
They further study how developers introduced these bugs, finding that 1) logic errors, such as missing cases or steps in the implementation of a routine and incorrect conditions of an if statement, dominate all studied errors; 2) algorithmic errors, including incorrect implementation of an algorithm and using an incorrect algorithm, rank second in terms of number; 3) some errors are language design errors, which are not introduced by developers but by designers and theorists. Park et al.~\cite{JEST} took into account this type of error when designing the fuzzer JEST for JavaScript engines.
In addition, though 89\% of all studied bugs could be fixed within 100 lines of code, it still took 186 days on average and a median of 24 days to fix the bug. This observation shows the difficulty in fixing front-end bugs.
Finally, they find that parameterized types, type argument inference, and parameterized class are the top three language features that can be utilized to trigger front-end bugs. In summary, this work paints a full picture of the front-end bugs of the four compilers and paves the way for designing new fuzzers for the compiler front-end and investigating front-end bugs of other popular programming languages.



\myparagraph{DL Compilers}
Besides bug studies for these traditional compilers, Shen et al.~\cite{qingchaoBugStudy} also make an effort to understand the symptoms, root causes, and distribution of the bugs in DL compilers.
They collect 603 bugs in three then-dominating DL compilers, including TVM ~\cite{TVM}, Glow~\cite{Glow}, and nGraph~\cite{nGraph}. Then, they categorize them into six symptoms and two root causes. They also attribute these bugs to three compilation stages.
Subsequently, they calculate the occurrence frequency of each root cause and each symptom, analyze the relationship between root causes and symptoms, and investigate the most bug-prone compilation stage.
The most exciting findings are as follows.
1) Type problem is the most common root cause, and tensor shape problem is the 3rd common. 
These data-related problems become the instigator of 32.5\% of the analyzed bugs. 
2) Crash is the most common symptom, accounting for 59.37\% of the total. 
Wrong Code is the second most common symptom (accounting for 25.04\%). These bugs manifest 
themselves as generating incorrect compiled DL models. Different from Crash, Wrong Code
are much less obvious and raises challenges in test oracle design.
3) The two most common root causes, type problems, and incorrect code logic, can induce all kinds of symptoms.
Therefore, to catch more bugs, all symptoms are worthy of attention.
4) High-level IR transformation is the most bug-prone compilation stage, accounting for 44.92\% of the total. 
This finding inspires Ma et al.~\cite{hirgen} to detect high-level optimization bugs in DL compilers.

Du et al.~\cite{DuBugstudy} also investigate bugs in DL compilers.
Specifically, they collect a total of 2717 bugs from TVM, Glow, nGraph, PlaidML, and TC, then
manually classify them by five bug types, including environment, compatibility, memory, document, and semantic.
They find that 
1) semantic and compatibility are two major bug types; 
2) more than one-third of bugs in DL compilers result in crashes or exceptions;
3) different types of bugs tend to have different symptoms.


These research efforts pave the way for the development of compiler fuzzing in two aspects. First, they demonstrate that compilers are still plagued by in-depth bugs, which necessitates and inspires compiler fuzzing. 
For instance, the bug study for Java-family compilers~\cite{jvmlanguagebugstudy} inspires Hephaestus~\cite{Hephaestus}, and the DL compiler study~\cite{qingchaoBugStudy} inspires HirGen~\cite{hirgen} to focus on testing the most bug-prone compilation stage.
Second, they list common bug symptoms, root causes, and bug-prone components or compilation stages, highlighting the fuzzing orientation. 
\section{Compiler Fuzzing}
\label{sec: Compiler Fuzzing}
Compiler fuzzing is a critical process that can detect compilers' defects. However, this task is not easy as it requires a large number of test programs to be constructed. 
The construction of these test programs is a challenging task: not only should the construction ensure the validity of the test programs (\ie the test programs should obey both the syntax rules and the semantics rule of the programming language), but also the constructed test programs should be diverse (\ie the test programs should contain as many language features as possible).
Additionally, merely relying on compiler crashes (\eg assertion failures and segmentation faults) to signal abnormal behavior may miss many in-depth bugs, thus limiting the bug-detection potential of the constructed test programs.
Therefore, it is necessary to develop new test oracles that provide a detailed assessment of whether a test program is successfully compiled.

In this section, we first outline the existing approaches for test program construction (Section \ref{sec:Test Program Construction Approaches}).
We also introduce the test oracle design (Section \ref{sec:testoracle}).

\subsection{Test Program Construction Approaches}
\label{sec:Test Program Construction Approaches}
\begin{table*}[ht]
  \centering
  
  \caption{
    A Summary of Papers about Test Program Construction
  }
  \resizebox{\linewidth}{!}{%
  \begin{tabular}{@{}c|c|c|c|c|c|c|c@{}}
  \toprule
  \multirow{4}{*}{\textbf{Paper}} 
   & \multicolumn{6}{c}{\textbf{Category}} \\ \cline{2-8}
   & \multicolumn{4}{c|}{\textbf{Generation-based}}  & \multicolumn{2}{c|}{\textbf{Transformation-based}} & \multirow{3}{*}{\textbf{Learning-based}}  \\  
   \cline{2-7}
   & \textbf{Grammar} & \textbf{User-Provided} & \textbf{IR} & \textbf{Auxiliary} & \textbf{Semantics} & \textbf{Semantics} &  \\
   & \textbf{Hardcoding} & \textbf{Specification} & \textbf{Design} & \textbf{Methods} & \textbf{Preserving} & \textbf{Changing} &  \\
  \midrule
  Eide and Regehr~\cite{Randprog} & $\checkmark$ & & & & & &\\
  Yang et al.~\cite{Csmith} & $\checkmark$ & & & & & &\\
  Even-Mendoza et al.~\cite{CsmithEdge} & $\checkmark$ & & & & & &\\
  Livinskii et al.~\cite{YARPGen} & $\checkmark$ & & & & & &\\ 
  Morisset et al.~\cite{C++memory} & $\checkmark$ & & & & & &\\
  Nagai et al.~\cite{Nagai2012RandomTO} & $\checkmark$ & & & & & & \\
  Liu et al.~\cite{NNSmith} & & $\checkmark$ & & & & & \\
  Chaliasos et al.~\cite{Hephaestus} & & & $\checkmark$ & & & & \\
  Groce et al.~\cite{swarmtesting} & & & & $\checkmark$ & & & \\
  Alipour et al~\cite{directedswarmtesting} & & & & $\checkmark$ & & & \\
  Chen et al.~\cite{HICOND} & & & & $\checkmark$ & & & \\ 
  Chen et al.~\cite{COTest} & & & & $\checkmark$ & & & \\ 
  Le et al.~\cite{EMI} & & & & & $\checkmark$ & & \\
  Le et al.~\cite{Athena} & & & & & $\checkmark$ & & \\
  Sun et al.~\cite{Hermes} & & & & & $\checkmark$ & & \\
  Lidbury et al.~\cite{manycore} &$\checkmark$ & & & & $\checkmark$ & & \\
  Jiang et al.~\cite{CUDAsmith} &$\checkmark$ & & & & $\checkmark$ & & \\
  Donaldson et al.~\cite{GLFuzz} & & & & & $\checkmark$ & & \\
  Donaldson et al.~\cite{spirv-fuzz} & & & & & $\checkmark$ & & \\
  Ma et al.~\cite{hirgen} &$\checkmark$ & & & & $\checkmark$ & & \\
  Chen et al.~\cite{classfuzz} & & & & & & $\checkmark$ & \\
  Chen et al.~\cite{classming} & & & & & & $\checkmark$ & \\
  Holler et al.~\cite{langfuzz} & & $\checkmark$ & & & & $\checkmark$ & \\
  Han et al.~\cite{CodeAlchemist} & & & & & & $\checkmark$ & \\
  Zhao et al.~\cite{JavaTailor} & & & & & & $\checkmark$ & \\
  Nagai et al.~\cite{Nagai2014} & & & & & & $\checkmark$ & \\
  Cummins et al. ~\cite{DeepSmith} & & & & & & & $\checkmark$ \\
  Liu et al.~\cite{DeepFuzz} & & & & & & & $\checkmark$ \\
  Xu et al.~\cite{DSmith} & & & & & & & $\checkmark$ \\
  Zhong~\cite{LeRe} & & & & & & & \\
  Chowdhury et al.~\cite{CyFuzz} & $\checkmark$ & & & & & &\\
  Chowdhury et al.~\cite{SLForge} & $\checkmark$ & & & & & &\\
  Chowdhury et al.~\cite{SLEMI} &  & & & & & $\checkmark$&\\
  Guo et al.~\cite{COMBAT} &  & & & & & $\checkmark$&\\
  Xiao et al.~\cite{MTXiao} &  & & & & & $\checkmark$&\\
  Zang et al.~\cite{JAttack}  & & & & & & & \\
  Liu et al.~\cite{liu2022coverage} &  & & & & & $\checkmark$&\\
  Luo et al.~\cite{GraphFuzzer} &$\checkmark$ & & & & & $\checkmark$ & \\
  Gro{\ss}~\cite{FuzzIL} & & & & & & $\checkmark$ & \\
  Gro{\ss} et al.~\cite{Fuzzilli} & & & $\checkmark$ & & & $\checkmark$ & \\
  Wang et al.~\cite{Skyfire} & & & & & & & $\checkmark$ \\
  Le et al.~\cite{Montage} & & & & & & & $\checkmark$ \\
  Park et al.~\cite{DIE}  &  & & & & & $\checkmark$&\\
  Tang et al.~\cite{DIPROM} &  & & & & & $\checkmark$&\\
  He et al.~\cite{SoFi} &  & & & & & $\checkmark$&\\
  Park et al.~\cite{JEST} &  & & & & & $\checkmark$&\\
  Wang et al.~\cite{FuzzJIT} &  & & & & & $\checkmark$&\\
  \bottomrule

  \end{tabular}
  
  }
\label{tab:ProgramConstruction}
\end{table*}
Compilers take programs as input and compile them to generate executables on different platforms.
This nature forces researchers to construct programs for compiler fuzzing.
However, program construction contains several challenges.
The main challenge is the promise of validity. The compilation is complicated and contains multiple stages.
A general compiler could be roughly divided into front-end, middle-end, and back-end. 
Take LLVM~\cite{LLVM} as example, 
The front-end handles lexical analysis, parsing, AST construction, and IR generation.
The middle-end works on LLVM IR implementation and related passes and optimizations.
The back-end plays the role of transforming LLVM IR into executables.
With large probability, invalid programs that do not obey the grammar of the programming language tend to fail at an early stage of lexical analysis or parsing.
They certainly cannot test the following stages.
Therefore, generating valid programs is a huge challenge.
Besides, to trigger as many bugs as possible, test programs should be expressive by intuition. 
In other words, programs should contain high-level language features, a complicated control flow, etc.
In previous works, test program construction strategies could be separated into three classes:
1) generation-based program construction, which follows the grammar of a subset of a programming language and generates a program automatically from scratch without reference to existing test programs,
2) transformation-based program construction, which mutates or synthesizes existing programs,
3) two-pronged program construction, which combines the above two construction strategies, and
4) learning-based program construction, which utilizes learning strategies and even DL models to learn the grammar of a programming language and generate test programs.
Table \ref{tab:ProgramConstruction} illustrates all collected papers and their categories.
We will elaborate on them by category in the following sections.

\subsubsection{Generation-based Program Construction}
\label{sec:generation-based}

Generation-based techniques focus on constructing test programs from scratch without any requirement for existing test programs.
In order to accomplish this objective, this category of techniques requires knowledge of the grammar of the target programming language.
In terms of knowing the grammar, these techniques
hardcode the grammar of the target language~\cite{Csmith, CsmithEdge, YARPGen}, 
or read user-provided specification about the grammar~\cite{langfuzz, NNSmith},
or contain an IR that could be translated into several downstream programming languages.
Besides these different generation strategies, researchers investigate effective configurations to direct program construction.
This section will introduce three different angles of test program generation and auxiliary methods for better generation.

\myparagraph{Program Generation through Grammar Hardcoding}
Yang et al.~\cite{Csmith} propose a generation-based fuzzer named Csmith for testing the C component of LLVM and GCC. This tool is based on Randprog~\cite{Randprog}.
Csmith generates C programs from scratch and utilizes these C programs to do differential testing by comparing the results of several executables produced by these compilers.
To prove validity, Csmith remembers the current generation point, finding all available and suitable variables, statements, and expressions. 
Besides, Csmith can avoid undefined behaviors (UB) in the C99 standard (\eg dereferencing a null pointer) to its best effort.
To filter out unsuitable candidates at one generation point and avoid undefined behaviors, Csmith hardcodes a subset of C grammar and obeys it to assure the validity of the generated program.
To enlarge the expressivity,
Csmith maintains a probability table containing all usable variables, statements, and expressions at the current program point and heuristically selects one candidate from the probability table. 
In practice, it can generate loop-rich, struct-rich, pointer-rich, and array-rich C programs with complicated control flows and a large amount of pointer dereference. 
As a then-effective compiler fuzzer, Csmith helped report over 325 previously unknown compiler bugs.
The success of Csmith has inspired many following works. 
CsmithEdge~\cite{CsmithEdge}, for instance, challenges Csmith's obsession with UB-free generation.
Specifically, CsmithEdge extends Csmith by probabilistically weakening the constraints to enforce UB-freedom.
In this way, CsmithEdge can possibly generate UB-contained programs.
It then employs off-the-shelf UB detection tools and dynamic analysis tools to detect the locations of UBs in the generated programs.
Finally, CsmithEdge can selectively remove all UBs and maintain two versions of programs: one with UB and one without UB.
The UB-free program can be used to detect miscompilation through differential testing.
And the non-UB-free program can be used to check if the compiler well-handles the UB (not crash or hang).
CsmithEdge proved its ability in bug detection by finding seven previously unknown miscompilation bugs in GCC, LLVM and Microsoft Visual Studio Compiler. In these seven bugs, five fixed ones could not be detected by Csmith.
Besides CsmithEdge, YARPGen~\cite{YARPGen} is another fuzzer for C/C++ compilers. 
In one testing campaign, YARPGen can conduct random generation and employ differential testing for bug detection, similar to Csmith.
But YARPGen contains generation policies to avoid saturation. In compiler fuzzing, a fuzzer tends to saturate (find no new bug) after being used for a while.
And reaching saturation does not mean the compiler is bug-free but means the fuzzer is biased in test program construction and cannot exercise specific components.
YARPGen attempts to resolve this problem by introducing generation policies: a mechanism that systematically skews the probability distribution and generates programs that are more likely to trigger certain optimizations.
Besides this innovation, YARPGen also pursues UB-free programs without resorting to the clumsy approach used by Csmith.
And YARPGen extends the language features by supporting generating C++ programs. With these contributions, it has detected more than 220 bugs in GCC, LLVM, and the Intel C++ compiler.

Besides these following works for fuzzing C/C++ compilers, Csmith also inspired other fuzzing works for a special compiler component or other compilers.
Morisset et al.~\cite{C++memory} add support for mutex, atomic variables, locking, and unlocking to Csmith.
In this way, this new version of Csmith can support thread-related language features to find concurrency bugs.
Nagai et al.~\cite{Nagai2012RandomTO} test the C component for optimizing arithmetic expressions by generating arithmetic expressions from scratch.
And the approach also involves heuristics to avoid UBs. 
After generation, the approach precomputes the result of each arithmetic expression and inserts runtime checks to compare the actual results and the precomputed actual results.
Besides these two approaches, Lidbury et al.~\cite{manycore} propose CLsmith based on Csmith. CLsmith can generate different types of OpenCL kernels in six modes.
Specifically, the implementation of CLsmith is an extension of Csmith with support for vector types and a rich set of vector operators in OpenCL.
Besides, CLsmith also absorbs EMI~\cite{EMI} and inserts dead codes, which will be introduced in section \ref{sec:transformation-based}.
Similar to CLsmith, CUDAsmith~\cite{CUDAsmith} fuzzes the CUDA C/C++ compiler with the same spirits: 1) random generation extended from Csmith, and 2) dead code injection inspired by EMI.
Also, Csmith inspires generation-based fuzzing for the Simulink compiler~\cite{CyFuzz, SLForge}. 
Recently, Csmith-like techniques have emerged for fuzzing the rapidly developed DL compilers.
Ma et al.~\cite{hirgen} propose HirGen to fuzz the most bug-prone compilation stage (\ie high-level optimization) of DL compilers.
HirGen can generate valid computational graphs by obeying built-in constraints extracted from DL compilers.
Besides, HirGen can also pursue the diversity of the graphs by following three coverage criteria. 
Furthermore, HirGen can convert one computational graph into several semantically equivalent high-level IRs to further test high-level optimizations.
Since HirGen also contains a transformation strategy, we will introduce it again in the transformation aspect in section \ref{sec:transformation-based}.

\myparagraph{Program Generation through Parsing User-Provided Specifications}
Hardcoding grammar is a safe strategy for ensuring valid program generation, but it is not flexible. 
Take Csmith as an example. The initial version avoids UBs under C99 standard, which is quickly outdated due to the rapid development of the C language.
Therefore, the developer of Csmith needs to update it every time the C standard is updated.
Besides, it requires a lot of engineering work to let Csmith support new language features of C and C++, not to mention supporting other languages.
Therefore, several existing program generation approaches take as input the user-provided specification about the grammar of a programming language. 
And they can conduct generation based on their understanding of these specifications.
Holler et al.~\cite{langfuzz} propose LangFuzz, which can read the grammar of different program languages (\eg PHP, JavaScript) and generate test programs of the target language.
Besides this innovation, LangFuzz also contains a transformation strategy that will be introduced in section \ref{sec:transformation-based}.
Liu et al.~\cite{NNSmith} propose NNSmith, a generation-based fuzzer for multiple DL compilers.
Specifically, NNSmith can generate valid computational graphs (\ie the test program for DL compilers) by reading user-provided specifications of constraints that must be satisfied.
And it can also generate valid inputs that will not trigger floating point exceptional values for the generated computational graph.

\myparagraph{Program Generation through IR Design}
Though flexible user-provided specification facilitates the generalizability of the program generation approach, it is still imperfect.
Writing specification requires understanding the grammar of at least a subset of one target language. 
And learning how to write specifications could not be a trivial task.
A feasible approach to freeing users from providing specifications while maintaining generalizability is to design an IR that can be translated into multiple real-life programming languages.
Chaliasos et al.~\cite{Hephaestus} propose Hephaestus, a generation-based fuzzer for the compilers of three JVM languages: Java, Kotlin, and Groovy.
To achieve this generalizability, Hephaestus involves a well-designed IR to abstract away differences among the target languages.
Since Hephaestus aims to detect type-related bugs, its IR syntax contains type-related features, such as class declaration with inheritance and type parameters.
During generation, Hephaestus can produce an IR and then translate it into one target language.
The utilization of IR design for better program construction is also used by transformation-based program construction~\cite{FuzzIL, Fuzzilli} (Section \ref{sec: Semantics-changing Transformation}).

\myparagraph{Better Program Generation through Auxiliary Methods}
The above program generation approaches all attempt to generate feature-rich test programs. 
But several researchers have observed that test programs should focus on expressing a small number of language features to test one or a few compiler optimizations and components rather than being feature-rich~\cite{swarmtesting, directedswarmtesting}.
Following this finding,
Swarm testing~\cite{swarmtesting} is proposed to restrict language features in test program generation.
Specifically, swarm testing enforces the avoidance of some features in one round of program construction but allows them in another round.
In this way, these features will not compete for space in each test or not obstruct the in-depth exploration of the possible bugs caused by one feature.
For instance, using "pop" for a stack data structure may prevent the fuzzer from detecting overflow-related bugs.
Based on swarm testing, Alipour et al~\cite{directedswarmtesting} further propose directed swarm testing, which tunes a set of available language features based on the analysis of the historical test programs.
Specifically, directed swarm testing can configure Csmith to cover a specific part of C compilers with a higher probability.
Besides, Chen et al.~\cite{HICOND} proposes HICOND to find a set of configurations (which regulates the involved language features in one construction) for better compiler fuzzing.
With the inference of configurations, HICOND can assist test program generators (\eg Csmith) to generate more diverse and bug-revealing test programs.
These configuration-based helpers facilitate the existing generators to generate bug-revealing test programs with a higher probability.
However, they only focus on configuring language features but not optimizations provided by the target compilers, which are also bug-prone~\cite{COTest}.
To address this issue, Chen et al.~\cite{COTest} propose COTest to explore the combinations of optimization flags given a test program generated
by Csmith.
Specifically, COTest first utilizes XGBoost~\cite{XGBoost} algorithm to predict the bug-triggering probability of a test program 
under an optimization flag combination.
Then, it uses a diversity augmentation strategy to select a set of diverse candidate combinations of optimization flags.
Finally, it selects the Top-K most possibly bug-triggering combinations.
As a helper for Csmith, COTest extends Csmith's ability in detecting bugs that are only detectable with the involvement of optimizations.

\subsubsection{Transformation-based Program Construction}
\label{sec:transformation-based}

Different from generation-based program construction, transformation-based program construction creates new test programs by performing a sequence of transformations on an existing program (\ie seed test program).
The transformation is implemented by either mutation (\eg statement insertion, statement deletion)~\cite{EMI, Athena, Hermes} or split plus splice (\ie split seed test programs into basic blocks and splice them together under constraints to form new test programs)~\cite{langfuzz, CodeAlchemist}.
Due to this nature, transformation-based program construction has a strong obsession with a certain characteristic of the constructed program (\eg the constructed program should behave the same as the seed program under some scenarios~\cite{EMI}, or the constructed program tends to contain bug-triggering features~\cite{langfuzz}).
And the implementation of transformation-based program construction can be slim in size (\eg Orion~\cite{EMI} has 500 lines of shell script and 1000 lines of C++ code, while Csmith~\cite{Csmith} contains 30-40K lines of C++ code).
To better showcase the basic ideas of the transformation-based program construction approaches and their evolution,
we divide these research efforts into two categories based on their transformation strategies: 1) semantics-preserving transformation, and 2) semantics-changing transformation.

\myparagraph{Semantics-preserving Transformation}
\label{sec:SPT}
Maintaining the semantics of the program after transformation is the gist of semantics-preserving transformation. 
Until now, a lot of semantics-preserving transformation approaches originate from the methodology named \textit{equivalence modulo inputs} (EMI)~\cite{EMI}. 
In short, EMI is a new concept of equivalence among a set of programs that behave the same after compilation given any input from an input set.
This equivalence contributes to test oracle design (Section \ref{sec:testoracle}) and also proposes challenges in program transformation:
how to transform the seed program without changing the semantics?

The first attempt is Orion~\cite{EMI}. It traverses the AST nodes of a program and stochastically prunes the unexecuted "dead" nodes.
In this way, the original program shrinks in size but still maintains the same semantics under the selected inputs.
Though Orion can generate equivalent test programs module inputs, its transformation strategy is too simple and blind.
To better the realization of EMI, Le et al.~\cite{Athena} propose Athena.
Besides code deletion, Athena also supports code insertion in the unexecuted code region.
Instead of this engineering effort, Athena also involves Markov Chain Monte Carlo (MCMC)
to sample the program space to generate diverse programs which differ from the seed program to a large extent.
Compared to Orion, Athena contains both code insertion and heuristics in generating new test programs.
But it still suffers from some limitations. First, because Athena can only manipulate dead code regions, the transformation is confined.
Second, compilers may eliminate dead code regions for optimization. In this case, the transformation is vain.
To solve this problem, Sun et al.~\cite{Hermes} propose Hermes, which can mutate the live code region.
Specifically, Hermes can synthesize a code snippet and insert it into a live code region to modify the control flow while maintaining the semantics of the program.
This transformation is achieved by, for example, inserting a code block whose conditional predicate evaluates to false.

The success of applying EMI to test C compilers invokes other research efforts for other compilers.
CLsmith~\cite{manycore}, CUDAsmith~\cite{CUDAsmith}, MT-DLComp~\cite{MTXiao}, SLEMI~\cite{SLEMI}, and COMBAT~\cite{COMBAT} all involve dead code injection to test the OpenCL C/C++ compiler, the CUDA C/C++ compiler,  DL compilers, and the Simulink compiler.
The basic thought of EMI is to construct equivalent programs modulo \textit{a set of inputs}, 
but this thought requires profiling the execution of the seed program.
To relieve this engineering burden, EMI can be further extended to construct equivalent programs modulo \textit{all inputs}.
For instance, GLFuzz contains 1) \textit{expression transformation} (rewrite a mathematical expression into its equivalent form) and 2) \textit{vectorization} (replace occurences of variables into the occurences of a vector of them).
These two transformation promises the generated programs are semantically equivalent to the seed program modulo all inputs.
Similar to GLFuzz, spirv-fuzz~\cite{spirv-fuzz} is also proposed to test graphics shader compilers.
In particular, spirv-fuzz targets at fuzzing the compiler of SPIR-V, an intermediate representation used by the Vulkan GPU programming model.
Similar to C and C++, SPIR-V also features undefined behaviors.
Spirv-fuzz contains a well-designed transformation strategy that can facilitate almost free test program reduction and test program deduplication.
Specifically, spirv-fuzz maximizes the independence among all transformations and favors simple transformations with a tendency toward atomicity.
In this way, it can further perform test program reduction and test program deduplication almost for free.
HirGen~\cite{hirgen} includes a transformation strategy to generate equivalent high-level IRs modulo all inputs.
Specifically, HirGen can reconstruct the function call chain of the IR by 
1) turning a global function F into the local ones inside other functions that call F,
2) wrapping a function F with an empty function which returns F and the corresponding call sites to F, and
3) inserting a helper function G to return a caller to function F and rewrite all other callers to F into callers to G.

\myparagraph{Semantics-changing Transformation}
\label{sec: Semantics-changing Transformation}
Different from semantics-preserving transformation, semantics-changing transformation transforms programs without keeping their semantics.
The benefit of this transformation strategy is that it can 
1) enrich the expressivity and diversity of the programs being transformed,  
and test specified components of the target compilers, which is the main benefit, and
2) prevent from generating error-contained test programs.

Chen et al.~\cite{classfuzz} propose classfuzz to gain the first benefit for fuzzing JVM implementations.
Specifically, classfuzz has several mutators (\ie the mutation operators, each of which implements one mutation strategy) and can
perform Markov Chain Monte Carlo (MCMC) method to select a sequence of mutators that have a large probability of triggering compiler bugs.
These mutators can modify the semantics of the seed program by, for instance, inserting/deleting methods into/from the existing classes.
However, classfuzz mainly produces illegal bytecode files to test the startup process of JVM. Since most of these illegal bytecode files
are rejected by the startup process, they are incapable of fuzzing the JVMs' bytecode verifiers and execution engines.
To achieve in-depth fuzzing JVM implementations, Chen et al.~\cite{classming} further propose classming. Specifically, classming can
1) mark the classfile's live bytecode and generate semantically different classfiles from the original by inserting/deleting/modifying instructions while obeying syntactic and structural constraints, and
2) select the constructed classfiles in each iteration using an acceptable choice for further classfile construction.
This way, classming can generate valid and diverse bytecode files and better test JVM implementations.

The idea of transforming bytecode is also applied to test JavaScript engines.
Gro{\ss} proposes FuzzIL to transform JavaScript bytecodes to generate more diverse ones.
Specifically, he designs an IR named \textit{IL}, which is more suitable for mutation.
He creates five mutators on this IR and utilizes them to construct new IRs from the old.
At the end of the transforming process, the new IRs are lowered into JavaScript bytecodes.
Similar to FuzzIL, Fuzzilli also mutates the IR program to generate new programs. But differently, Fuzzilli is designed to test the JIT compiler of the JavaScript engine.
Not all fuzzing efforts on JavaScript engines are conducted on self-designed IRs.
Park et al.~\cite{DIE} find that some properties of the test programs can help trigger bugs in JavaScript engines. 
So while mutating the existing test programs, they recognize and retain these properties. 
The focus on some special bug-triggering properties is also applied by fuzzing work for C compilers. Tang et al.~\cite{DIPROM} propose DIPROM, which can mutate the existing test programs to make them contain warning-sensitive structures. This way, the constructed test programs are probable to detect compiler warning defects.
The mere mutation may introduce syntax and semantics errors. Take JavaScript as an example. Its dynamic type system increases the difficulty in generating valid test programs at both syntax level and semantics level, especially semantics level~\cite{SoFi}. To address this issue, He et al.~\cite{SoFi} propose SoFi to mutate the existing JavaScript test programs with fine-grained analyses. Besides, SoFi can repair the semantics error based on the error information. To improve the diversity of the constructed test programs, SoFi incorporates a dynamic reflection mechanism to detect novel attributes and methods of an object that are not present in the seed test programs. These novel attributes and methods will be inserted into the seed test programs to increase semantics diversity.

Regarding the testing for DL compilers and DL inference engines, Liu et al.~\cite{liu2022coverage} propose TZER to mutate the low-level IR of TVM. They propose two kinds of mutation strategies. One is general-purpose mutations, including inserting, deleting, and replacing low-level nodes in the low-level IR. The other is domain-specific mutations, including nesting existing loops, rewriting memory-related operators, and thread-binding for diversifying the parallel computation flows.  
This way, TZER can construct low-level IRs that can trigger more low-level optimizations, and thus more low-level optimization bugs will be dug out.
Luo et al.~\cite{GraphFuzzer} propose to mutate the generated DL models with six mutation strategies. In this way, their approach can further improve the degree of diversity of their generated models, and thus detect more bugs in DL inference engines.

Different from the above-mentioned techniques that transform test programs by mutations, 
LangFuzz~\cite{langfuzz} splits a test program into code fragments and splices them to implement program transformation.
Specifically, LangFuzz can parse bug-triggered test programs given the grammar and replace a code fragment with another code fragment (generated from scratch or extracted from another test program) of the same type.
After transformation, LangFuzz can recombine several possibly bug-revealing code fragments together and may trigger new bugs. CodeAlchemist~\cite{CodeAlchemist} adopts a similar approach to splicing code fragments for fuzzing JavaScript engines. And to tackle the difficulty in generating syntax-valid and semantics-valid JavaScript test programs, CodeAlchemist conducts type analysis and def-use analysis to verify if two code fragments can be spliced. For instance, it checks whether a variable is defined before being used and whether several types match for conducting the operations. JEST~\cite{JEST} does not break test programs for JavaScript into code fragments but directly splices the initial small seed test programs and then mutates them to increase the semantics diversity.
The fuzzing approach of utilizing historical bug-revealing test programs is also applied by a recent research effort named JavaTailor~\cite{JavaTailor} for fuzzing JVM implementations.
Specifically, JavaTailor can extract the ingredients from historically bug-revealing test programs and synthesize these ingredients randomly to create valid classfiles.
To achieve validity, JavaTailor can reuse variables in the synthesized classfile or construct new definitions to fix the broken syntactic and semantic constraints.
Similarly, the implementation of FuzzJIT~\cite{FuzzJIT} also involves the observation of bug-triggering features in the historical bugs of the JIT compilers in JavaScript engines. The developers of FuzzJIT notice that arrays, objects, and special numbers are bug-relevant. So they hardcode them in the mutation strategies of FuzzJIT to let it generate bug-revealing test programs.

As for the second benefit, Nagai et al.~\cite{Nagai2014} propose some transformation strategies on the basis of their previous work~\cite{Nagai2012RandomTO}
to eliminate undefined behaviors. Specifically, their previous work avoids generating long arithmetic expressions because they are more susceptible to produce UBs.
In their new work~\cite{Nagai2014}, they can mutate the UB-contained expressions into UB-free ones by inserting a new operation, flipping an operation, etc. In this manner, their approach can produce lengthy arithmetic expressions without concern for UBs.

\subsubsection{Two-pronged Program Construction}

Modern compiler fuzzing approaches tend to gain benefits from both program generation and program transformation.
Their strategies for combining these two program construction methods can be divided into two categories:
1) generation-then-transformation combination, and
2) generation-for-transformation combination.
CLsmith~\cite{manycore}, CUDAsmith~\cite{CUDAsmith}, HirGen~\cite{hirgen}, GraphFuzzer~\cite{GraphFuzzer}, and Fuzzilli~\cite{Fuzzilli} belong to the first category. 
The first step of these techniques is generating test programs.
As for the second step, CLsmith and CUDAsmith adopt the idea of EMI~\cite{EMI} to generate equivalent test programs modulo a set of inputs (Section \ref{sec:EMI-equivalence}). HirGen transforms the generated high-level IRs (\ie the test programs for DL compilers) into the equivalent modulo all inputs (Section \ref{sec:Non-EMI-equivalence}).
These transformations address the test oracle problem (Section \ref{sec:testoracle}).
Different from the above three techniques GraphFuzzer~\cite{GraphFuzzer} implements six mutation strategies to further diversify the generated DL models. 
Specifically, GraphFuzzer includes four model-level mutations, including graph edges addition, graph edges removal, block
nodes addition, and block nodes removal; and two source-level
mutations, including tensor shape mutation and parameter~.
mutation. These mutation aims to change the semantics (Section \ref{sec: Semantics-changing Transformation}) of the DL models, and thus achieve higher diversity.
Similarly, Fuzzilli generates an IR, mutates it to achieve diversity, and finally lowers it into JavaScript programs, to test JavaScript JIT compilers.

LangFuzz~\cite{langfuzz} is in the second category. Its main idea is to recombine the code fragments in the historical bug-revealing test programs to form new test programs.
But since the substitutable code fragments of some special type are inadequate, LangFuzz can generate code fragments from scratch to enrich these materials.

\subsubsection{Learning-based Program Construction}

Wang et al.~\cite{Skyfire} propose Skyfire to learn the syntax and semantics rules from existing test programs and generate valid test programs with the knowledge. A similar learning idea is also adopted by TVMfuzz, a prototype for fuzzing TVM proposed by Shen et al.~\cite{qingchaoBugStudy}. TVMfuzz can learn the API call chains from the existing test programs for TVM and generate new and valid test programs.

With the development of DL models in the natural language processing (NLP) field, several researchers have investigated the use of these models in generating valid test programs to test compilers.
The first attempt was made by Cummins et al.~\cite{DeepSmith}. They utilized Long Short-Term Memory (LSTM) architecture of Recurrent Neural Network to learn over one million lines of valid OpenCL code mined from GitHub.
As for test oracles, build crash and build timeout are caught and reported immediately.
Other outcomes of test cases require differential testing. 
Specifically, when some testbeds behave the same while others deviate, the same behavior of the majority of testbeds is chosen as the correct behavior.
And abnormal behavior of other testbeds will be recognized as a potential bug as reported.
This technique has proven its ability in bug detection by reporting 67 bugs in 1000-hour testing.
And the test cases this technique generates are two orders of magnitude smaller than the state-of-the-art on average
and require more than 3.0X less time to generate.
Similarly,
Lee et al.~\cite{Montage} applies LSTM models to learn the relationships among the AST fragments from the ASTs of the existing JavaScript test programs. This way, their technique, Montage, can generate new test programs conforming to syntax rules and semantics rules.
Besides these efforts in fuzzing OpenCL C/C++ compiler and JavaScript engines, Liu et al.~\cite{DeepFuzz} utilize a generative sequence-to-sequence model to generate well-formed C programs.
Their technique, DeepFuzz, improves the line, function and branch coverage
And it helps detect eight confirmed bugs of GCC.
Furthermore, Xu et al.~\cite{DSmith} observes that then-current deep-learning-based compiler fuzzers tend to generate syntax errors because these fuzzers fail to capture long-distance dependencies of syntax.
Therefore, they propose a new technique, DSmith, to solve this problem.
Specifically, DSmith introduces a LSTM unit and an attention mechanism to memorize hidden states of all tokens and capture their interactions.
With this design, DSmith can increase the parsing pass rate by an average of 19\% and significantly improve the code coverage of the compiler.

\subsubsection{Other Approaches}

Not all test program constructions are of the above four categories.
For instance, 
with the observation that the test program of a compiler is reusable to test others,
Zhong~\cite{LeRe} proposes a technique named LeRe to construct test programs by extracting from bug reports of both GCC and Clang and utilize 
them to test other compilers.
Specifically, LeRe can parse the attachment and analyze the descriptions from a bug report. 
In parsing the attachment,
LeRe can determine whether the attached file is a code file by analyzing the file name.
In analyzing the descriptions,
LeRe can extract the test program, the error message and the comment separately.
Finally, LeRe performs differential testing using these test programs to detect previously unknown compiler bugs.

Zang et al~\cite{JAttack} propose JAttack, a framework that enables developers to incorporate their domain knowledge on fuzzing compilers. Specifically, developers need to write a template program, a special Java program with many holes to be filled. JAttack can fill these holes with suitable and randomly selected 
expressions or values to generate valid Java test programs. 
Besides utilizing developers' observation and witness in fuzzing, JAttack can also extract templates
from existing Java projects to gain some implicit knowledge for better fuzzing.

\subsection{Test Oracle Design}
\label{sec:testoracle}

In software testing, a test oracle is a mechanism for determining whether a test has passed or failed.
Similar to general software testing, compiler fuzzing requires well-designed test oracles to 
detect in-depth compiler bugs.
Research efforts about the test oracle design for compiler fuzzing can be divided into two categories:
1) differential testing~\cite{differentialtesting} and 2) metamorphic testing~\cite{metamorphictesting}.
In short, differential testing cross-checks the compilation results of a test program across multiple \textit{compilation tracks} while
metamorphic testing cross-checks the compilation results of several equivalent test programs concerning a single compiler.

\subsubsection{Differential Testing}

\begin{table*}[ht]
  \centering
  \caption{
    Usage of the Selected Compilation Tracks
  }
  \scriptsize
  \begin{tabular}{@{}c|c@{}}
  \toprule
  \textbf{Paper} & \textbf{Role of the Collected Compilation Track(s)} \\
  \midrule
  Sheridan et al.~\cite{differentialtesting} & Referee \\
  Kitaura et al.~\cite{10.1145/3278186.3278192} & Referee \\
  Liu et al.~\cite{liu2022coverage} & Oracle \\
  Ofenbeck et al.~\cite{RandIR} & Oracle \\
  Morisset et al.~\cite{C++memory} & Oracle \\
  Le et al.~\cite{Proteus} & Oracle \\
  Bera et al.~\cite{bera:hal-01299371} & Oracle \\
  Bernhard et al.~\cite{JIT-Picking} & Oracle \\
  Wang et al.~\cite{FuzzJIT} & Oracle \\
  Yang et al.~\cite{Csmith} & Voter \\
  Lidbury et al.~\cite{manycore} & Voter \\
  Jiang et al.~\cite{CUDAsmith} & Voter \\
  Even-Mendoza et al.~\cite{CsmithEdge} & Voter \\
  Livinskii et al.~\cite{YARPGen} & Voter \\
  Hawblitzel et al.~\cite{10.1145/2491411.2491442} & Voter \\
  Chen et al.~\cite{classfuzz} & Voter \\
  Sun et al.~\cite{10.1145/2884781.2884879} & Voter \\
  Barany~\cite{10.1145/3178372.3179521} & Voter \\
  Chen et al.~\cite{COTest} & Voter \\
  Zhong~\cite{LeRe} & Voter \\
  Ma et al.~\cite{hirgen} & Voter \\
  Chowdhury et al.~\cite{CyFuzz} & Voter \\
  Chowdhury et al.~\cite{SLForge} & Voter \\
  Chowdhury et al.~\cite{SLEMI} & Voter \\
  Guo et al.~\cite{COMBAT} & Voter \\
  Tang et al.~\cite{DIPROM} & Voter \\
  Park et al.~\cite{JEST} & Voter \\
  \bottomrule

  \end{tabular}
  
\label{tab:differentialtesting}
\end{table*}

Differential testing was first proposed by McKeeman et al.~\cite{differentialtesting} as a complement to regression testing.
In compiler fuzzing, differential testing requires multiple comparable compilation tracks for compiling a single test program
and warns of the existence of possible bugs in these compilation tracks if the compilation results are inconsistent.
Besides using different compilers for one programming language (\eg LLVM and GCC for C/C++) and different versions of one compiler,
turning on different optimizations for compiling a single test program and comparing the results under these optimizations
is also considered differential testing in compiler fuzzing~\cite{JunjieSurvey}.
Concretely, a compilation track is determined by a tuple of the compiler, the version, the optimization, 
and other configurable plugins provided by the compiler. 
Modifying any element of this tuple will cause a different compilation track and provide possibly 
different compilation results (e.g, 
different IRs, different assemblies or different execution results).
For instance, as for the differential testing for C compilers, two comparable compilation tracks could be
$GCC(version:\ 12.3 \oplus optimization:\ O2)$ \textit{v.s.} $LLVM(version:\ 13.0 \oplus optimization:\ O2)$.
Differential testing on multiple compilation tracks can be divided into the following three categories
based on the selected compilation tracks.
In general, the selected compilation tracks can work as 
1) the referee for the compiler under test, 
2) the oracle that always performs correct compilation, or
3) voter for the correct compilation.
Table \ref{tab:differentialtesting} details the categories of the differential testing approaches proposed in the collected papers.

\myparagraph{Referee Compilation Track}
Sheridan~\cite{Sheridan_2007} propose a differential testing approach for a C99 compiler, named PalmSource
Cobalt ARM C/C++ embedded cross-compiler (referred to as PalmSource hereafter).
Specifically, Sheridan selects another two compilers, the GNU C Compiler in C99 mode (version 3.3), 
and the ARM ADS 1.2 assembler, as the referee compilers. These two compilers can offer two compilation tracks as referees.
After comparing the outputs of the executables generated by these compilation tracks given the same inputs,
353 bugs of PalmSource have been found.
There are several insights underlying this differential testing:
1) In the referee compilation track selection, a reasonable level of quality of the referee compilers is required for efficient bug detection for the target compiler.
2) There is no need to promise that the referee compilers and the target compiler accept the same language.
A substantial overlapping among them is fine.
3) Bugs in the target compiler are likely to result in output after compilation that differs from
the outputs of the referee compilation tracks for the same input. And the high quality of the referee
compilers implies the large probability of the existence of potential bugs in the target compiler.
Besides selecting other compilers to construct referee compilation tracks, Kitaura et al.~\cite{10.1145/3278186.3278192}
propose to select the older versions of the compiler for building compilation tracks and use them to detect performance
bugs of the latest version. Since the older versions is more stable and well-tested, it is high-quality and can work as the referee.

\myparagraph{Oracle Compilation Track}
Besides selecting reference compilers of high quality, 
researchers also select the oracle compiler that always provides the correct compilation
to judge the correctness of the compiler under test.
Ofenbeck et al.~\cite{RandIR} propose RandIR to test the embedded compiler of a domain special language (DSL)
for the multi-stage programming (MLP) in Scala. 
In short, MLP aims at rewriting an operation (\eg an arithmetic operation) 
into the symbolic counterpart to hide details.
In this way, dead code elimination and code motion can be easily conducted without knowing too many operation details.
After that, the embedded compiler compiles the optimized symbolic counterpart to the vanilla Scala functions.
RandIR allows the users to provide a grammar for the operations and it can compose these operations into a DSL (an IR in this context).
Then it can conduct differential testing with inquiry to an oracle compiler.
The oracle compiler is implemented by a direct generation of the vanilla Scala function from the given grammar.
RandIR can also utilize another embedded DSL compiler as the reference compiler to conduct differential testing.

A similar testing idea is applied to the optimizations for C11/C++11 memory model~\cite{C++memory}.
Specifically, Morisset et al. select the compilation track excluding optimizations as the oracle.
They compare the memory trace
of the compiled code under this oracle compilation track
with that under the compilation tracks including optimizations and select out all mismatches
and report all potential bugs included.
The optimization-excluding compilation track must not contain optimization-related bugs, and thus
is a representative oracle.
Le et al.~\cite{Proteus} propose Proteus with a similar oracle compilation track selection strategy to test
link-time optimizers (LTO).
Specifically, Proteus selects the compilation track excluding LTO as the oracle.
They compare the compilation result of the oracle and the compilation results of all LTO-including compilation tracks.
Any mismatch may imply the existence of a bug.
The strategy of selecting the optimization-excluding compilation track as the oracle is also applied by another research effort
~\cite{bera:hal-01299371}, which tests the deoptimization of JIT compilers by comparing optimization-excluding compilation tracks and
deoptimization compilation tracks. The optimization-excluding compilation track must not contain deoptimization-related bugs,
so it is an oracle.

Similar to traditional compilers, DL compilers also provide optimizations.
Liu et al.~\cite{liu2022coverage} use optimization-excluding compilation track as the oracle to test TVM.
In particular, they create two test oracles that go beyond detecting crashes, using the following principles: 1) a low-level intermediate representation (IR), whether optimized or not, should yield identical outcomes, and 2) low-level optimizations should not result in degraded performance.
Similar to the previous work, the optimization-excluding compilation track is recognized as the oracle since it must not have optimization-related bugs.

JavaScript engines also contain JIT optimizations.
Specifically, they may turn on the JIT compilation during interpreting JavaScript bytecodes to compile frequently exercised code paths and then execute them to save time costs. JIT compilers, in this context, work as an optimizer for the interpreter. Bernhard et al.~\cite{JIT-Picking} propose JIT-Picking to detect silent (\ie the bugs implicitly induce errors without crashing the compilation process) and in-depth bugs inside this optimizer. 
Specifically, JIT-Picking extends Fuzzilli by executing the constructed JavaScript programs twice: once with the JIT compiler enabled and once without it. The compilation track without involving the JIT compiler must not contain JIT compiler bugs. So it works as an oracle to check if the JIT-enabled compilation track contains silent bugs. A similar idea is adopted by FuzzJIT~\cite{FuzzJIT}. FuzzJIT can insert some optimization-triggering statements (\eg multiple calls to a function with the exact same arguments) into the seed test program. It can further monitor whether the optimizations performed by the JIT compiler of JavaScript engines corrupt the result. For instance, it can record the result of the first function call (the first call would not enable JIT optimizations) and the result of the last function call (at this call, the JIT optimizations have been invoked). In this test oracle, the optimization-disabled compilation track is the oracle for the optimization-enabled compilation track.

\myparagraph{Voter Compilation Track}
In reality, high-quality references and always-correct oracles may not be available.
In this situation, researchers require adequate compilation tracks and let them vote for the correct compilation results.
Csmith~\cite{Csmith} utilizes this strategy. 
In short, Csmith collects $LLVM(version:\ 1.9-2.8)$, $GCC(version:\ 3.[0-4].0)$, and $GCC(version:\ 4.[0-5].0)$
and uses all of them to compile a single test program.
Each compilation track votes for its compilation result and the majority is recognized as the possible correct one.
Though there is no ground truth that the majority tends to be correct. 
But in the experiment, Csmith never finds a split vote without an obvious consensus, nor finds any cases that the majority agree on the same incorrect result. 
The following works~\cite{manycore, CUDAsmith, CsmithEdge, YARPGen} also follow this voting mechanism in differential testing.
Similarly, Hawblitzel et al.~\cite{10.1145/2491411.2491442} propose to collect voter compilation tracks by using 1) different
versions of a compiler, 2) different optimization settings of a compiler, 3) different architectures (\eg ARM and X86) and
4) different compilation modes (\eg Just-In-Time and Machine Dependent Intermediate Language). Each compilation track produces
an assembly language output, a vote for the correct compilation.
To strictly detect bugs in any compilation track, Hawblitzel et al. compare each vote by proving the equivalence of these
assembly language programs.
Specifically, they use a symbolic differencing tool SymDiff, a program verifier Boogie and an automated
theorem prover Z3. 
The assembly language programs are first converted into the Boogie language with the help of SymDiff.
Next, they use Boogie to generate verification conditions. Finally, they use Z3 to prove the validity of these conditions.
Any nonequivalence between any pair of compilation tracks is a witness of a possible bug in one of the compilation tracks.
In differential testing on JVM implementations, Chen et al.~\cite{classfuzz} propose classfuzz to generate
class files for testing. During differential testing, classfuzz collects several JVM implementations
and several versions of one JVM implementation as voters. 

Besides the implementations, differential testing can also detect bugs in the specification. 
Park et al.~\cite{JEST} propose a differential testing framework for both JavaScript engines and the JavaScript specification.
They select four JavaScript engines, including Google V8, GraalJS, QuickJS, and Moddable XS. And after constructing test programs, they execute them on these engines. They argue that if only a minority of the selected engines report an error, those engines may contain bugs. However, if the majority of the engines report an error, it suggests that the specification itself may contain defects, as all the engines are implementations of the same specification.

Differential testing is also applied to the diagnostics component of compilers. Sun et al.~\cite{10.1145/2884781.2884879} propose
to detect warning defects of C compilers by differential testing.
Specifically, they collect compilation tracks by selecting 1) several compilers (GCC and Clang), 2) different versions of a compiler,
and 3) multiple optimization levels (\eg -O1, -Os, -O2, -O3).
After obtaining the warning messages, they design more than 200 parsers to parse them and compare the parsing results.
Tang et al.~\cite{DIPROM} propose DIPROM to generate test programs containing warning-sensitive structures.
Their selection of compilation tracks is identical to that of Sun et al.~\cite{10.1145/2884781.2884879}.
Besides diagnostics component testing, Barany~\cite{10.1145/3178372.3179521} and Kitaura et al.~\cite{10.1145/3278186.3278192} collect
voter compilation tracks to detect performance-related bugs by differential testing; Chen et al.~\cite{COTest} collect voter compilation
tracks using multiple optimization levels to detect optimization-related bugs.

The idea of differential testing is also applied to DL compiler fuzzing~\cite{hirgen, NNSmith} and Simulink compiler fuzzing~\cite{CyFuzz, SLForge, SLEMI, COMBAT}.
In short, DL compilers utilize low-level compilers (\eg LLVM) and low-level platforms (\eg CUDA, OpenCL) to generate executable codes.
HirGen~\cite{hirgen} utilizes this feature and collects multiple compilation tracks that only differ in the usage of the low-level compiler/platform.
And NNSmith~\cite{NNSmith} uses different DL compilers to collect compilation tracks.
As for the Simulink compiler, it provides multiple compilation options including simulation modes and optimization levels. 
CyFuzz~\cite{CyFuzz}, SLForge~\cite{SLForge}, SLEMI~\cite{SLEMI} and COMBAT~\cite{COMBAT} collect compilation tracks with different options.

\subsubsection{Metamorphic Testing}

\begin{table*}[ht]
  \centering
  \caption{
   Metamorphic Testing for Compilers 
  }
  \scriptsize
  \begin{tabular}{@{}c|c@{}}
  \toprule
  \textbf{Prerequisite of Equivalence Relations} & \textbf{Papers} \\
  \midrule
  \multirow{8}{*}{EMI-equivalence}
    & Le et al.~\cite{EMI} \\
    & Le et al.~\cite{Athena} \\
    & Sun et al.~\cite{Hermes} \\
    & Lidbury et al.~\cite{manycore} \\
    & Jiang et al.~\cite{CUDAsmith} \\
    & Chowdhury et al.~\cite{SLEMI} \\
    & Guo et al.~\cite{COMBAT} \\
    & Xiao et al.~\cite{MTXiao} \\
  \midrule
  \multirow{5}{*}{Non-EMI-equivalence}
   & Tao et al.~\cite{Mettoc} \\
   & Donaldson and Lascu~\cite{7811329} \\
   & Donaldson et al.~\cite{GLFuzz} \\
   & Ma et al.~\cite{hirgen}  \\
   & Xiao et al.~\cite{MTXiao} \\ 
  \bottomrule

  \end{tabular}
  
\label{tab:metamorphictesting}
\end{table*}

The core idea of the solution to the test oracle problem of metamorphic testing~\cite{metamorphictesting}
is to construct deformation relations.
In short, metamorphic relations specify how the changes to the input would impact the output.
For instance, suppose that a program is an implementation of the sine function.
Given an input $x$, the program should output $sin(x)$.
During testing, it is difficult to determine the output of an arbitrary input, for instance, $1$,
if we do not have a correct calculator for the sine function.
However, the equation $sin(x) = sin(\pi - x)$ can help construct a metamorphic relation:
given any input $x$, if we change it into $\pi - x$, the output of the program should not change.
We can use this metamorphic relation to determine the expected output of $\pi - x$ and compare it with the actual output.
As for constructing metamorphic relations for compilers, the most widely adopted method is constructing equivalence relations between test programs.
Based on the prerequisite of the equivalence, we divide all collected papers into two categories:
1) EMI-equivalence and 2) non-EMI-equivalence.
The first category adopts EMI~\cite{EMI} methodology to transform the seed program. 
The equivalence is modulo a set of inputs and does not always hold.
Differently,
the second category proposes some equivalence relations without any prerequisite.
The collected papers and their details are shown in Table \ref{tab:metamorphictesting} and the following sections.

\myparagraph{EMI-equivalence}
\label{sec:EMI-equivalence}
Le et al.~\cite{EMI} introduce a concept named Equivalence Modulo Inputs (EMI) for testing compilers.
In short, the key insight behind EMI is to produce seemingly different, but actually equivalent test programs on a set of inputs. 
Besides introducing EMI, this work also provides an EMI implementation, Orion, to detect bugs in LLVM and GCC.
Orion can transform the seed test program by randomly deleting non-executing code snippets on a set of inputs.
Athena~\cite{Athena} and Hermes~\cite{Hermes} are two upgraded version of Orion.
In short, Athena can both insert and delete statements into and from the non-executed code snippet.
Hermes can further mutate the executed code snippet to its semantically equivalent variants under a set of inputs.
These EMI-family techniques contain different semantic-preserving transformation strategies for constructing new test programs.
The implicit metamorphic relation underlying these EMI-family techniques is that
changing the seed test program into the constructed ones will not change the compilation result (pass or fail).
And it does not change the output of the compiled code given any input from the input set.
The EMI-family techniques also include CLsmith~\cite{manycore}, CUDAsmith~\cite{CUDAsmith}, 
SLEMI~\cite{SLEMI}, COMBAT~\cite{COMBAT} and MT-DLComp~\cite{MTXiao}.
Though these testing techniques target at different compilers, they all require the profiling phrase
to acquire the runtime information to generate semantically-equivalent test programs \textit{w.r.t.} a set of inputs.
For instance, MT-DLComp~\cite{MTXiao} records the outputs of all nodes during profiling and uses these outputs $Os$
to construct a sequence of operators whose final output based on $Os$ is $0$.

\myparagraph{Non-EMI-equivalence}
\label{sec:Non-EMI-equivalence}
Different from EMI-equivalence, non-EMI-equivalence is not built on the input.
In other words, the non-EMI-equivalence relationship holds regardless of the input.
For instance, Tao et al.~\cite{Mettoc} propose Mettoc, a metamorphic testing tool
for compilers.
Mettoc can generate equivalent test programs by
1) constructing equivalent expressions (\eg $e_1 + e_2 = e_2 + e_1$, $e*2 = e+e$),
2) constructing equivalent compounded statements consisting of a sequence of assignments (\eg swap two assignment statements without modifying the value),
and
3) constructing equivalent compounded
statements which may include loop and branch structures (\eg eliminate unused variables).
After generating equivalent test programs, Mettoc
further uses the compiler under test to compile them into executables.
Inconsistent outputs imply that the compiler contains a bug.

Donaldson and Lascu~\cite{7811329} propose a metamorphic testing approach for graphics shader compilers.
They analyze the opaque values (fresh variables that will take fixed values at runtime but whose values are unknown to the compiler) and use them to
compose an expression that evaluates to false. 
Then, they use the expression as the conditional of an if statement whose body cannot be
executed due to the expression must be false.
Since the expression contains an opaque value, the compiler cannot optimize away the expression.
Therefore, this code insertion is always successful.
Besides the above dead code injection, Donaldson et al.~\cite{GLFuzz} also introduce five other program
transformation strategies to generate non-EMI-equivalent test programs for testing graphics shader compilers.
These transformation strategies are dead jump injection, live code injection, expression mutation, vectorization, and control flow wrapping.
In short, the dead jump injection injects a jump (break, return, continue, or discard statement) as the body of an if statement whose conditional
is an always-false expression.
The live code injection injects actually-executed code which does not change the semantics of the original program.
In practice, the inserted code is drawn from another program and the variables contained are all renamed to avoid name clashes.
To achieve semantical preservation, the injected code accesses disjoint data from the original program.
The expression mutation rewrites a boolean or numerical expression into its equivalent form (\eg $e = e + 0$).
The vectorization rewrites the declarations of multiple variables into a vector declaration. 
And it rewrites all occurrences of these variables
into the accesses to the member of the vector.
The control flow wrapping complicates the control flow while keeping the original execution times of each statement. This way,
this transformation is semantics-preserving.

The idea of constructing non-EMI-equivalence is also applied in DL compiler testing.
HirGen~\cite{hirgen} constructs equivalent high-level IRs by rewriting function call chains (Section \ref{sec:SPT}).
Besides constructing EMI-equivalents, 
MT-DLComp~\cite{MTXiao} constructs equivalent DL models by inserting always-yield-zero operators
into the seed models.

\section{Open Challenges and Research Opportunities}
\label{sec: Open Challenges and Research Opportunities}

Despite the considerable advancements made in compiler fuzzing, there are still challenges that need to be addressed. This section aims to identify some of these challenges that may be tackled in future research.

\myparagraph{Diversity of Test Programs}
Generating valid test programs is not a major hurdle in contemporary compiler fuzzing, as most published research papers offer well-designed algorithms to ensure program validity.
However, the diversity of test programs is still under exploration.
First, there are no definitive criteria for quantifying diversity.
Certain researchers argue that obtaining higher code coverage indicates that test programs are diverse~\cite{liu2022coverage}, as it is intuitive to assume that including more language features will cover a greater range of compiler code related to these features.
Some researchers introduce some coverage criteria to measure directly the diversity of test programs~\cite{GraphFuzzer, hirgen}. They suggest that the test programs should contain certain possibly bug-triggering elements and advocate for monitoring the construction process to encourage the test programs to include them.
Some research works~\cite{Athena, Hermes, classming} are a continuation of their predecessors~\cite{EMI, classfuzz}. These follow-up works enhance mutation strategies, resulting in higher diversity compared to their predecessors.

Though different works adopt different strategies to increase diversity, there is no systematic study on it.
Besides, large language models (LLMs) have proven their ability to synthesize simple DL models to some degree~\cite{deng2023large, deng2023large2}.
Despite their effectiveness, utilizing such strategies to construct more complex DL models or even type-rich and valid programs written in modern languages, such as C++ or Java, remains challenging and warrants further exploration.

\myparagraph{Extensibility and Generalizability}
The current successful compiler fuzzing tools, ranging from generation-based ~\cite{Csmith, YARPGen} to transformation-based~\cite{EMI, Athena, Hermes}, all hardcode their supported grammar or mutation strategies internally, resulting in the fact that scaling these tools is not easy.
Further development based on them requires a comprehensive understanding of their source codes, which is time-consuming~\cite{DeepSmith}.
Designs that are easily extensible are necessary and require further exploration.

In addition, most of the collected papers propose a compiler fuzzer for specific components of the compiler(s) for only one programming language or DSL.
For instance, Csmith~\cite{Csmith} fuzzes the C components of LLVM and GCC.
classfuzz~\cite{classfuzz} and classming~\cite{classming} fuzzes the JVM implementations.
HirGen~\cite{hirgen} and TZER~\cite{liu2022coverage} only fuzz TVM.
While their approaches are generalizable, transferring them to other compilers can be labor-intensive~\cite{manycore, CUDAsmith, DeepSmith}.
To address this issue, some research efforts have been made.
Some works focus on using DL models to learn the grammar of the programming language, to reduce labor costs~\cite{DeepSmith, DeepFuzz}.
Certain works propose an approach to read user-provided grammar specifications and then generate the programs of the target programming language after comprehending the grammar~\cite{langfuzz}.
LLMs have the potential to mitigate labor costs as well. However, the development of prompts to fully unleash their ability for program construction and synthesis is still being explored~\cite{deng2023large, deng2023large2}.

\myparagraph{Evaluation for Compiler Fuzzing}
Though there exist research efforts offering guidance on evaluating fuzzing works~\cite{evaluatefuzz}, a rigorous academic discussion on how to evaluate compiler testing techniques is still required due to the following reasons.
Establishing a consensus on a uniform evaluation approach or building benchmarks for each compiler is crucial for evaluating the efficacy of novel compiler fuzzing techniques.
\section{Conclusion}
\label{sec: Conclusion}

Compiler fuzzing plays a crucial role in improving the correctness of compilers, which are fundamental to software development.
In this survey, we systematically studied high-quality 58 papers on the topic of fuzzing modern compilers and compiler-like tools that are still under active development and maintenance. 
This survey primarily focuses on the two main challenges for effective and efficient compiler fuzzing: test program construction and test oracle design. In terms of test program construction, we examine how current research efforts construct valid and diverse test programs. Regarding test oracle design, we discuss how innovative test oracles are designed to detect in-depth bugs.

Despite the significant advancements made in compiler fuzzing, there are still several outstanding issues that require attention. It is our hope that this survey will attract newcomers to the field of compiler fuzzing and encourage experts to continue making progress in addressing the existing challenges.
\printbibliography

\end{document}